# Role of redox additive modified electrolytes in making Na-ion supercapacitors a competitive energy storage device

Sudipta Biswas[a,+], Debabrata Mandal[b], Ananya Chowdhury[a] and Amreesh Chandra[a,b,*]

[a]Department of Physics, [b]School of Nano Science and Technology,

Indian Institute of Technology Kharagpur, Kharagpur-721302.

Email: sudiptabiswas058@gmail.com[+], *achandra@phy.iitkgp.ac.in*[*]

**Abstract**

The study shows the importance of moving towards hollow nanostructures for obtaining next-generation supercapacitors and batteries. Amongst various sodium-based electrode materials, $NaFePO_4$ is one of the most promising cathodic material because of its high electrochemical potential, structural, and thermal stability. To use this material in high performance Na-ion based energy storage device, it is imperative to combine it with a suitably optimized carbon structures and a redox additive modified electrolyte. This strategy is unequivocally established in the paper. On addition of redox additive, the performance can be improved by 50%. This makes the Na-ion supercapacitor competitive with other metal ion based systems. the performance by as high as 50%. This makes the Na-ion supercapacitors competitive with other metal ion based systems.

## 1. Introduction

The future of Na-ion based energy devices depends on the development of the smarter nanostructures, which can combine with other complementary components like electrolytes, counter electrode, etc. and deliver enhanced performance. These materials will have to handle the chemistry of bigger ions like $Na^+$. There are a few materials, which have now been obtained and are being used in Na-ion supercapacitors or batteries[1-3]. These include Na-based oxides, phosphates, binary, ternary, oxides solid solutions and composite[4-12]. More recently, the strategy of using pseudo 2-dimentional structures of Na-ion electrode have been proposed[13, 14]. Although these have led to appreciable improvement in the performance of supercapacitors, still much more is desired[15-17]. This means that additional strateges have to be proposed to make Na-ion supercapacitor competitive.

Use of redox additive or electrolyte modifiers has been successfully used in many energy storage devices[18-21]. But, such strategy remains largely ignored for Na-ion based supercapacitors. The additives help in increasing the redox reactions while ensuring reduction in the resistance for ion transfer.[22] Therefore, simultaneous enhancement in the capacity and power/enrgy density is observed.[21] In this paper, we propose and establish a successful procedure of using a battery type $NaFePO_4$ nanostructures and rGO combination for developing competitive Na-ion supercapacitors. Further, in combination with sodium persulphate (NPS) modified electrolyte, these supercapacitors can show further enhancement of 30% increase in the performance. The value thus achievable by Na-ion supercapacitors becomes competitive with other established supercapacitors[23-25]. To make further improvement, the choice of carbon based negative electrode is also optimised and described in the paper. The device is able to deliver a specific capacitance of 195 F g$^{-1}$ at a current density 1 A g$^{-1}$, with a high cycling stability of more than 85% at 3 A g$^{-1}$ specific current and a high energy density of 35 kW kg$^{-1}$ with a specific power 889 Wh kg$^{-1}$.

## 2. Experimental

### 2.1 Material used

Ferric nitrate nonahydrate (Fe(NO$_3$)$_3$, 9H$_2$O), stearic acid (C$_{18}$H$_{36}$O$_2$), ammonium dihydrogen phosphate ((NH$_4$)H$_2$PO$_4$), tri-sodium citrate di-hydrate (Na$_3$C$_6$H$_5$O$_7$, 2H$_2$O), sodium persulfate (Na$_2$S$_2$O$_8$) and sodium hydroxide (NaOH) were procured from LobaChemie Pvt. Ltd. (India) and MerckSpecialities Pvt. Ltd. (India), respectively. All the analytical grade precursors were used directly without further purification.

*NaFePO$_4$*: Porous and hollow NaFePO$_4$ microstructures were synthesized using a one-pot facile hydrothermal route followed by calcination in air. In a typical experimental procedure, 25 ml of 0.1 M ferric nitrate solution was mixed with 25 ml of 0.1 M stearic acid solution. Subsequently, 245.1 mg trisodium citrate (Na$_3$C$_6$H$_5$O$_7$, 2H$_2$O) was added and the solution was stirred for 2 h. An appropriate quantity of ammonium dihydrogen phosphate ((NH$_4$)H$_2$PO$_4$) was added to the precursor solution to obtain Na:Fe:PO$_4$ concentration ratio of 1:1:1. 50 ml of this yellow-colored solution was then transferred to Teflon-lined stainless steel (capacity 250 ml) autoclave. The autoclave was kept at 180 ºC for 24 h, before allowing it to slowly cool down to room temperature. The precipitate was collected by centrifugation at 3200 rpm. The precipitate was washed three times using de-ionized water before being vacuum dried at 70 ºC for 12 h. The dried sample was crushed and annealed at 600 ºC for 4 h in air to obtain NaFePO$_4$ powder.

*Carbonaceous material*: Activated carbon was purchased from Merck Industries Pvt. Ltd. and used without any treatment. GO was synthesized using a modified Hummer's methods. Initially, graphite powder was added into sulfuric acid (98%) in an ice bath. Potassium permanganate was gradually added into this mixture. Following stirring for 10 min, sodium nitrate was added. This solution was further stirred for 3 h at 35 °C before dilution by deionized (DI) water. The reaction was terminated by the addition of $H_2O_2$ (30%). The obtained light yellow color graphite oxide was washed with dilute HCl (volume ratio of 1:10 for HCl to DI water) to remove the residual ions. Finally, the suspension was filtered and washed with DI water several times till a pH~7 was obtained. The brown-colored aqueous graphite oxide suspension was sonicated for 40 min for exfoliation. Finally, the sonicated suspension was centrifuged at 3500 rpm for 30 min. Subsequently, GO was reduced by $NaBH_4$. In a typical reduction, GO (0.6 gm) was stirred in 200 ml pure water and was dispersed finely in water in ultrasonic bath for 3 h. Then, while the GO suspension was being stirred, $NaBH_4$ (2 gm) was added to the GO suspension. The solution was heated in an oil bath at 80-100 °C under a water-cooled condenser or without the condenser for 12 and 24 h, respectively. The water level was kept constant during all the reductions. The precipitated rGO samples were washed with acetone, DI water and ethanol. Finally, the washed rGOs were dried in an oven under vacuum.

### 2.3 Material characterization

The phase formation of the synthesized materials was confirmed by analyzing the powder x-ray diffraction (XRD) profiles, using Rigaku miniflex 600 diffractometer utilizing Cu-Kα (λ = 0.15406 nm) as the excitation wavelength, in 2θ range 15-60°. Scanning electron microscopy (SEM CARL ZEISS SUPRA 40) was used for the morphological analysis. The Brunauer-Emmett-Teller (BET) surface area and pore size were estimated by analyzing the $N_2$ adsorption-desorption isotherms obtained by a Quantachrome Nova Touch surface area and pore size analyzer. Zeta potential and particle size of the samples were measured using a Horiba Scientific Nano Particle Analyzer SZ-100. For the electrochemical characterization of the synthesized materials and fabricated devices, cyclic voltammetry (CV) and galvanostatic charge-discharge (GCD) was performed using the MetrohmAutolab (PGSTAT302N) potentiostat-galvanostat and Zahner zennium pro electrochemical test station.

### 2.4 Electrode preparation and electrochemical characterization

*3-Electrode measurement*: Electrochemical measurements were performed using graphite sheet as the current collector. This was coated with slurries containing electrode materials. Slurries were prepared using an active material, polyvinylidene fluoride-co-hexafluoropropylene (PVDF-HFP) binder and activated carbon using acetone as the mixing media. The ratio of the materials was 8:1:1 for active material, PVDF and activated carbon, respectively. The obtained slurries were drop casted onto a graphite sheet (dimension 1 cm × 1 cm) and vacuum dried at 80 °C for 12 h. The measurements were performed in three electrode configurations using a 2 M aqueous NaOH electrolyte. The mass of the electrode was kept at ~1 mg. Platinum wire and Ag/AgCl/3.0 M KCl were used as the counter and reference electrodes, respectively. For the study with redox additive in the electrolyte, various concentration (0 to 120 mM) of NPS were added to the 2 M pristine electrolyte.

*Assymetric device*: To construct the device CR2032 type cells were used, which had the width and thickness of 20 and 3.2 mm, respectively. Whatman glass fiber was used as the separator for the device. The details of device fabrication, mass loading etc. are given in S.I.

### 3. Result and discussion

#### 3.1. Physiochemical characterizations

For an electrochemically active material, parameters such as phase, particle morphology/ size, along with the nature of pores and pore-size/ volume, and surface area are detrimental factors in deciding the final electrochemical response. A representative XRD is shown in Fig. 1 (a). Single-phase formation of $NaFePO_4$ could be confirmed by comparing the XRD data with the JCPDS card no. 04-012-9665, corresponding to the *Pnmb* space group. X-ray photoelectron spectrometry (XPS) studies, to confirm the chemical configurations of $NaFePO_4$ powders is given in S.I.

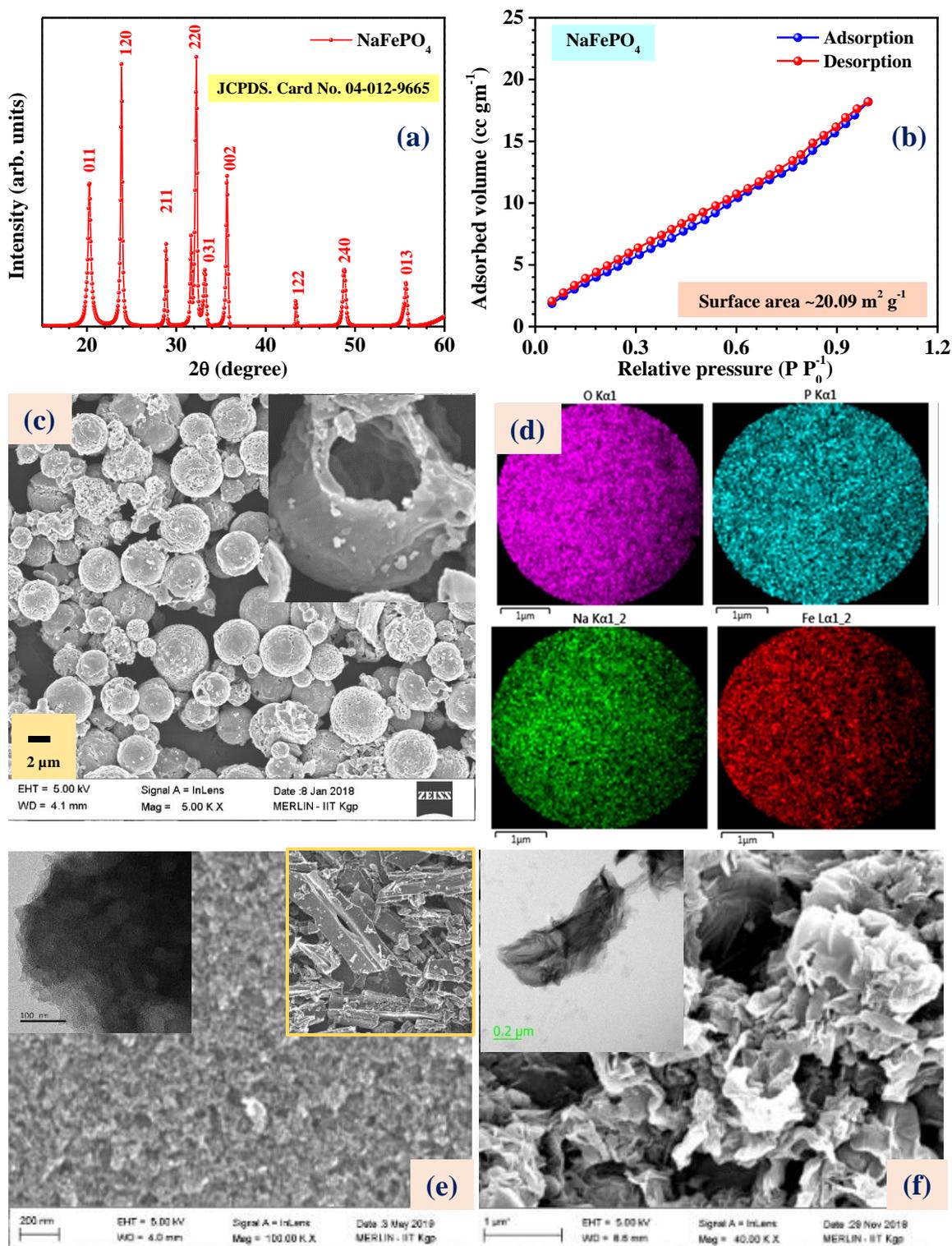

**Fig. 1** (a) X-ray diffraction pattern, (b) BET N$_2$ adsorption-desorption isotherm, (c) SEM micrograph (inset: zoomed surface of the sphere) and (d) elemtal mapping of the NaFePO$_4$ structures, SEM micrograph (inset: TEM) of (e) activated carbon and (f) rGO, respectively.

Along with phase, the active surface area, shape and volume of the pores also are critical for optimizing the material's performance. The surface area, pore size distribution, particle size distribution, and zeta potential values are shown in Fig. 1(b) and Fig. S1(a-c). The BET adsorption-desorption curves showed Type IV isotherms, indicating slit-shaped mesopores of ~ 6 nm. BET surface area was found to be ~ 20 $m^2 g^{-1}$. Zeta potential of a material denotes the electro-kinetic potential in colloidal dispersions.

Zeta potential is the potential difference between the dispersion medium and the stationary layer of fluid attached to the dispersed particle. During any chemical reaction, electro-positivity of the surface is determined by the overall pH of the solution. When a material gets dispersed in any liquid, the oppositely charged ions ($OH^-$) are separated and proceed towards the surface of the dispersed particles to form a layer, that is called as stern layer. On the other hand, the released $H^+$ ions reduces the pH of the solution and zeta potential shifts towards the positive regime. The NFP particles returned higher electro-positivity, which can only happen when there is an increased volume to facilitate $OH^-$ accommodation. This phenomenon helps to accommodate ions towards the surface and hence increases the capacitance of the material. The surface charge (zeta potential) was obtained as -41 mV, as shown in Fig. S1(b) for $NaFePO_4$ nanostructures. The large value of zeta potential indicated towards positively charged surface, which can accommodate more charges and also enhance the specific capacitance of the material.

The morphology of the particles has a dominant role in deciding the final electrochemical characteristics. It determines the overall surface area and available paths for ions adsorption and desorption. Fig. 1(c) shows the SEM micrographs of $NaFePO_4$. The size of the microspheres varied in the range 1-3 µm, with distinctly stabilized cavity in the middle. Recently, hollow nanostructures of metal oxides have been suggested as electrode materials for next-generation supercapacitors. Additionally, such structures have advantages like higher surface area, larger pore size, and transport channels in comparison to their solid counterparts due to porous nature. The TEM micrographs for the materials are shown in Fig. S2(a). The associated elemental mapping (shown in Fig. 1(d)) and EDAX (shown in Fig. S2(b)) data for the NaFePO$_4$ microspheres confirmed the atomic ratio of Na:Fe:P:O to be 1.05: 1: 1.08: 3.57, which was the desired ratio.

Carbon based materials were also characterized and the phase formation of the carbon materials was confirmed by analyzing the XRD profiles. Fig. S4 shows the X-ray diffraction pattern of activated carbon and rGO. The peak near 25° is the characteristic of (002) plane of graphite. Fig. 1(e,f) shows the SEM micrographs of activeated carbon and the TEM image showed the porous nature of the activated carbon and rGO, respectively. Fig. S5(a) shows the N$_2$ adsorption-desorption isotherms for activated carbon and reduced graphene oxide, respectively. The estimated surface area for the activated carbon and rGO were 933 m$^2$ g$^{-1}$ and 83 m$^2$ g$^{-1}$, respectively. Corresponding pore size distribution is shown in Fig. S5(b). The FTIR and Raman spectra of both the carbon structures are shown in Fig. S7(a) and S7(b), respectively. Particle size distribution of the activated carbon and rGO as seen from Fig. S7(a), showed, average particle size was ~0.5 μm and 3 nm, respectively. The corresponding Zeta potential were -6.61 and 4.43 mV, for activated carbon and reduced graphene oxide, repectively, as shown in Fig. S7(b).

### 3.2 Electrochemical characterization of hollow and porous structures

The stable potential window for NFP electrodes were found to be -0.3 to 0.5 V in 2 M NaOH electrolyte. Cyclic voltametry curves, with increasing scan rates, are shown in Fig. 2 (a). The maximum specific capacitance, at 5 mV s$^{-1}$ scan rate, was ~125 F g$^{-1}$. As the scan rate was increased to 200 mV g$^{-1}$, the values decreased to 31 F g$^{-1}$. Therefore, the capacitance retention was found to be ~25%, when the scan rate was increased by 40 times. From the CD profiles, shown in Fig. 2(b), the maximum capacitance was found to be 142 F g$^{-1}$ in NaOH electrolyte, at 1 A g$^{-1}$ current density. The under utilization of bulk capacitance at high specific current stands behind the decrease of specific capacitance. At higher specific currents, the transfer of electrons towards the electrode is faster and hence the increase of voltage is rapid. Consequently, the electrode has less time to stay at a certain voltage causing low charge transfer, lowering in specific capacitance value is observed. The specific capacitance values $NaFePO_4$ based electrode material, in 2 M NaOH electrolyte, at different scan rates and current densities, are tabulated in Table 1.

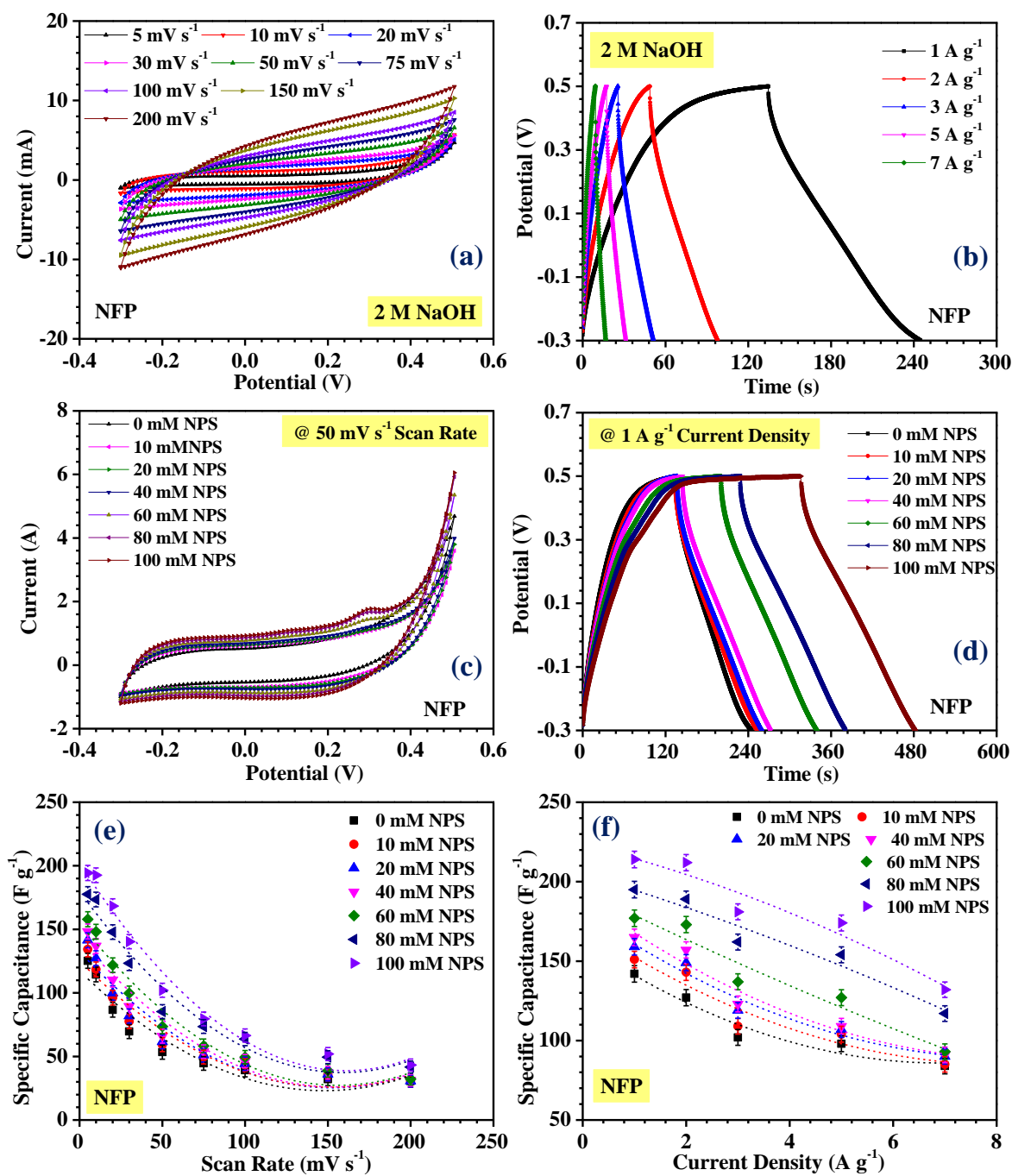

**Fig. 2** (a) CV, (b) CD profiles of NFP in 2 M NaOH, variation of (c) CV and (d) CD with increasing redox concentration (NPS) in the electrolyte, specific capacitance variation of NFP with (e) scan rate and (f) current density in pure and redox modified electrolytes.

The charge transport kinetics of the electrode was studied by electrochemical impedance spectroscopy (EIS) analysis. Nyquist plots give information about electrode-electrolyte interactions and equivalent series resistance (ESR). The ESR values for hollow NFP microsphere was ~1.36 and 2.23 Ω, before and after cycling test. Cyclic stability of the electrode is an important property to determine its commercial use. Hence, the cycling stability of the material was tested and ~93.6 % of capacity retention was observed after 1000 cycles at 3 A g$^{-1}$ current desnsity, as depicted in Fig. 3(d).

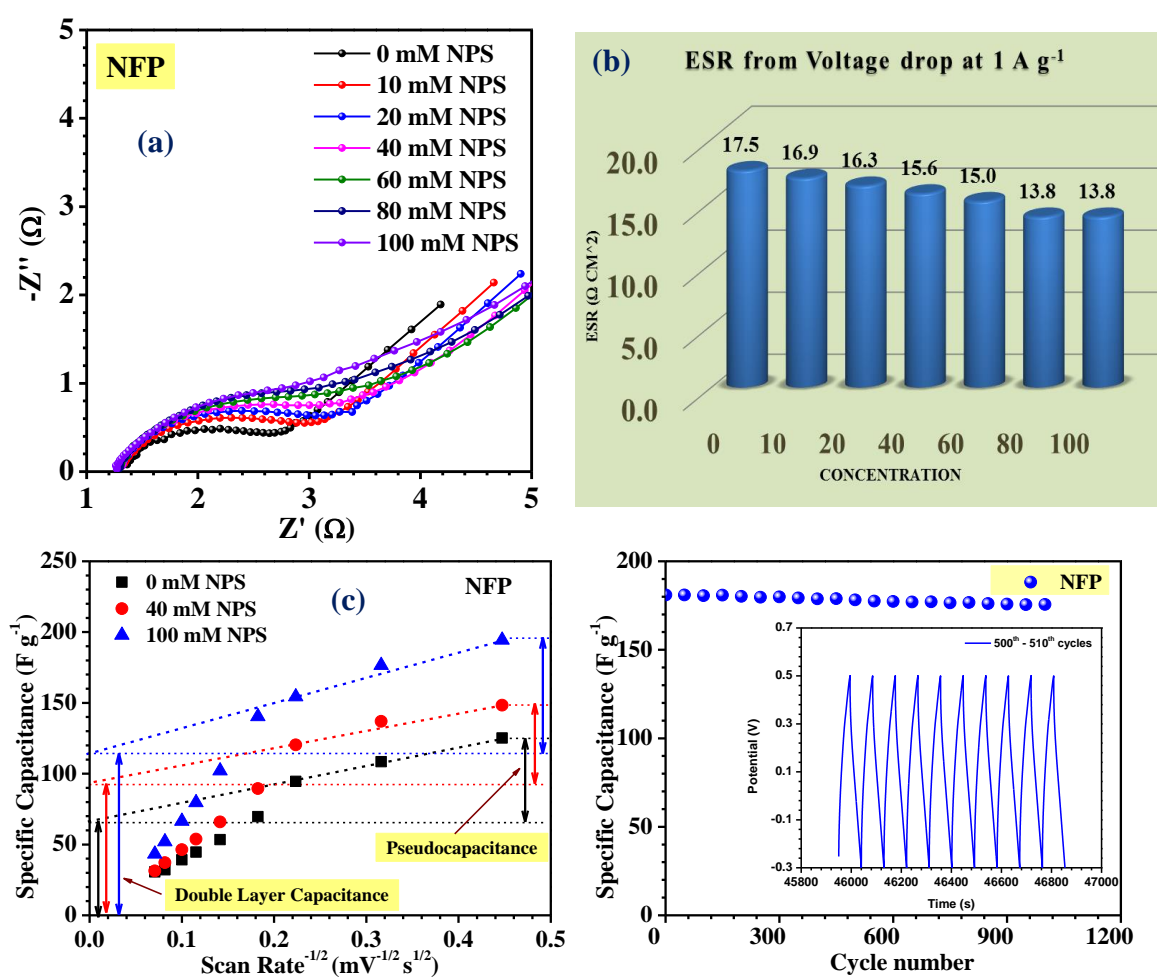

**Fig. 3** (a) Nyquist plot, (b) estimation of ESR from the CD profiles, (c) quantification plots and (d) cycling stability of NFP in pure and redox modified electrolytes.

Following optimization of electrode with the pristine electrolyte, the values for specific capacitance was needed to be further improved, if the capacitor has to become competetive. The next option available was to use the strategy of modifying the electrolyte. A suitable redox addititve i.e. sodium persulphate was chosen to modify the redox performance of NaOH. The concentration of the redox additive was varied from 10 to 100 mM. Initially, very little change was observed. But, on increasing the concentration, enhancement in specific capacitance was appreciable. The improvement, with increasing redox concentration, can be seen from the CV and CD profiles shown in Fig. 2(c, d), respectively. At 10 mM concentration of NPS, a specific capacitance, at 5 mV s$^{-1}$, was ~134 F g$^{-1}$. The specific capacitance value from CD was estimated as 151 F g$^{-1}$, at 1 A g$^{-1}$ discharge current density. The value at different current densities and increasing redox concentrations are shown in Table 1. The variation of specific capacitance with scan rate and current density for pure as well as redox modified electrolyte is shown in Fig. 2 (e, f), respectively. It can be seen that, at 10% redox additive concentration, 7% increment in specific capacitance was observed. So, for optimizion of the additive concentration, more redox additive was introduced to the pristine 2 M NaOH electrolyte. The specific capacitance values increased upto certain values before showing water splitting. This can be clearly seen from Table 1. Quatification curves also showed increment in the charge transfer, which is reflected on the enhanced contribution from the redox reaction induced increment in the pseudocapacitance. Charge discharge study also showed a similar trend. At 1 A g$^{-1}$ discharge current density, maximum specific capacitance delivered by NaFePO$_4$ was 214 F g$^{-1}$ at 100 mM redox added electroyte. The value decreased at higher concentrations.

**Table 1.** Specific capacitance of NaFePO$_4$ electrode with increasing increasing current density in 2 M NaOH and 2 M NaOH + increasing concentration of NPS.

| Current density (A g$^{-1}$) | Specific Capacitance (F g$^{-1}$) after redox (NPS) addition | | | | | | |
|---|---|---|---|---|---|---|---|
| | 0 mM | 10 mM | 20 mM | 40 mM | 60 mM | 80 mM | 100 mM |
| 1 | 142 | 151 | 159 | 165 | 177 | 195 | 214 |
| 2 | 127 | 143 | 149 | 157 | 173 | 189 | 212 |
| 3 | 102 | 109 | 119 | 123 | 137 | 162 | 181 |
| 5 | 98 | 104 | 107 | 109 | 127 | 154 | 174 |
| 7 | 84 | 85 | 90 | 91 | 93 | 117 | 132 |

The possible equation for the redox reactions by the electrolytes could be written as the following:

$$Na_2S_2O_3 \leftrightarrow 2Na^+ + S_2O_8^{2-}$$

$$Fe^{2+} + S_2O_8^{2-} \leftrightarrow 2Fe^{3+} + SO_4^{-*} + SO_4^{2-}$$

Introduction of redox additive would improve the ionic conductivity of the electrolyte as well as provide higher number of solvated ions in the electrolyte. This leads to additional path ways for electron transfer, thus resulting in performance improvement. This could be confirmed by the electrochemical impedance spectroscopy (EIS) measurement. The ESR value of the electrode decreased with the increase in the concentration of the redox additive, indicating improved conductivity of the electrolyte and resulting in higher charge storage. For 2 M NaOH, the ESR value was estimated as 1.36 Ω. As concentration of redox additive was increased to 5%, ESR value reduced to 1.27 Ω. The nature of the curve changed towards the low-frequency region of the Nyquist plot, the nature of supercapacitor showed deviation from expected trend. Thus it is clear that the strategy of using redox additive was resulting in synergistic effect and can make Na-ion supercapacitor competitive.

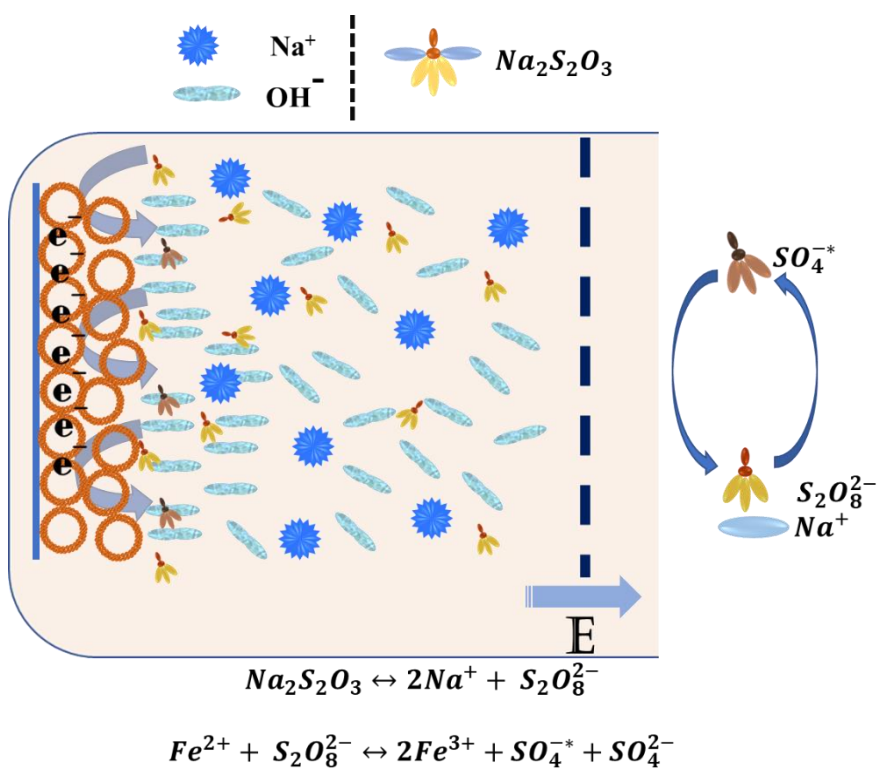

**Scheme 1** Sematic representation of the mechanism for redox modified supercapacitor.

To improve the performance further, choice of complimentary carbon electrode is essential. This has been reported recently for Na-ion supercapacitors[17]. Therefore, standard activated carbon and rGO were choosen as two types of negative electrode for the NaFePO$_4$ based device. The activated carbon was first tested in 2 M NaOH electrolyte in the potential window of -1 V to 0 V. The corresponding CV profile is shown in Fig. 4(a). Without addition of the NPS, activated carbon delivered maximum 144 F g$^{-1}$ at 1 A g$^{-1}$. For, 100 mM NPS modified electrolyte, the specific capacitance inceased slightly, as shown in Fig. 4(e). Hence, the electrochemical behaviour of the activated carbon was stable in redox modified electrolyte, which can be predicted from the EDL behaviour of the carbon. Actvated carbon showed <10% increment in the specific capacitance after redox incorporation. So, it can be concluded that redox additive mainy affects the pseudocapacitive materials.

As the rGO structures have higher EDL and redox behaviour with respect to the activated carbon, these structures were also tested with the addition of redox additive in the electrolyte. Fig. 4(c, d) depict the CV profiles for rGO in pure and 100 mM NPS modified NaOH electrolytes, respectively. Maximum specific capacitance for the rGO was ~188 F g$^{-1}$ at 5 mV s$^{-1}$ scan rate and 196 F g$^{-1}$ at 1 A g$^{-1}$ current density. Higher performance of this material as negative electrode can be attributed to higher surface area and higher porosity of these 2D structures. There are functional groups attached to the sheet like structures, which help to give additional redox behaviour to the material. Additionally, there are many oxygen functional groups present. Amongst these, the carboxyl and phenol actively take part in the Faradaic redox reactions when placed in the alkaline electrolyte with hydroxyl ions (OH$^-$). So, rGO shows both EDLC as well as pseudocapacitance. The corresponding charge transfer equations can be written as follows:

$$-COOH + OH^- \leftrightarrow -COO^- + H_2O + e^-$$

$$\equiv C-OH + OH^- \leftrightarrow \equiv C-O^- + H_2O + e^-$$

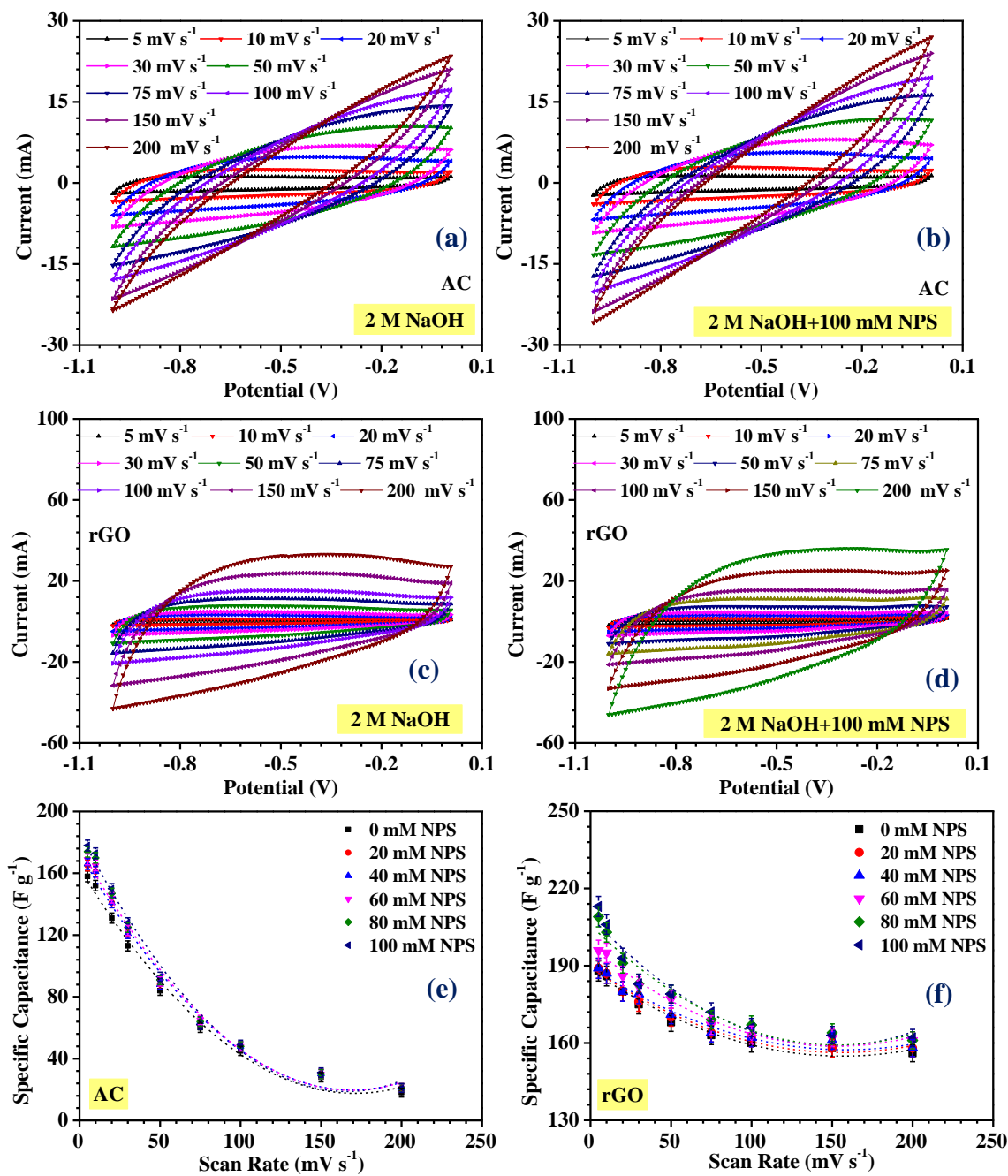

**Fig. 4** CV profiles of activated carbon and rGO in (a, c) 2 M NaOH, (b, d) 100 mM NPS added electrolyte, variation of specific capacitance with different redox concebntration for (e) activated carbon and (f) rGO.

These reactions are enhanced by the addition of the redox additive in the electrolyte. The specific capacitance of the rGO increases with the addition of NPS in the 2 M NaOH based electrolyte. Specific capacitance of rGO electrode at 2 M NaOH + 100 mM NPS electrolyte was found to be 215 F g$^{-1}$ at 5 mV s$^{-1}$ scan rate. So, we observed ~16% increament in specific capacitance for rGO, as the charge collection increased in the case of rGO. The variation of specific capacitance for rGO based electrode in both pure and redox modified electrolytes is depicted in Fig. 4(f).

Three electrode study, with and without redox additive, show the effect of the electrolyte on the electrode performance. To construct asymmetric device, mass balancing is essential. Following the mass balancing process, the desired mass ratio of positive and negative electrode was estimated as ~1.6, 0.8, 1.1 and 1.2 at 5 mV s$^{-1}$ scan rate in pure and redox modified electrolytes, with AC and rGO electrode, respectively. Fig. 5(a) depicts the CV profiles of NFP//AC device at different scan rates, from 5 to 200 mV s$^{-1}$. The device showed an optimized working potential window of 1.3 V (0-1.3 V). CV profiles of the device at each scan rate showed nearly rectangular shape, indicating dominant contribution from EDLC type behaviour. The galvanostatic charge-discharge measurements, performed at various current densities, are shown in Fig. 5(c). The charge-discharge profiles showed nearly linear and symmetric behaviour, exhibiting high Coulombic efficiency of ~95%. Fig. 5(b, d) depicts the CV and CD profiles of NFP//AC device at scan rates from 5 to 200 mV s$^{-1}$, using 2 M NaOH + 100 mM NPS electrolyte. With the addition of the redox additive electrolyte the voltage window of the device remained unaltered but the curve showed distortions. This can be attributed to the modified redox activity within the electrolyte. Maximum specific capacitance delivered by the NFP//AC device, in pure electrolyte, was 23 F g$^{-1}$, which increased to 41 F g$^{-1}$ after the addition of 100 mM NPS to the electrolyte. This meant nearly 78% increment in the specific capacitance. At 1 A g$^{-1}$, current density, the device capacitance increased from 23 F g$^{-1}$ to 36 F g$^{-1}$. So, redox modified electrode showed ~1.5 fold increment in the specific capacitance for NFP//AC combination. The specific capacitance variations of NFP//AC device, with scan rate and current density, are shown in Fig. 5(e, f), respectively.

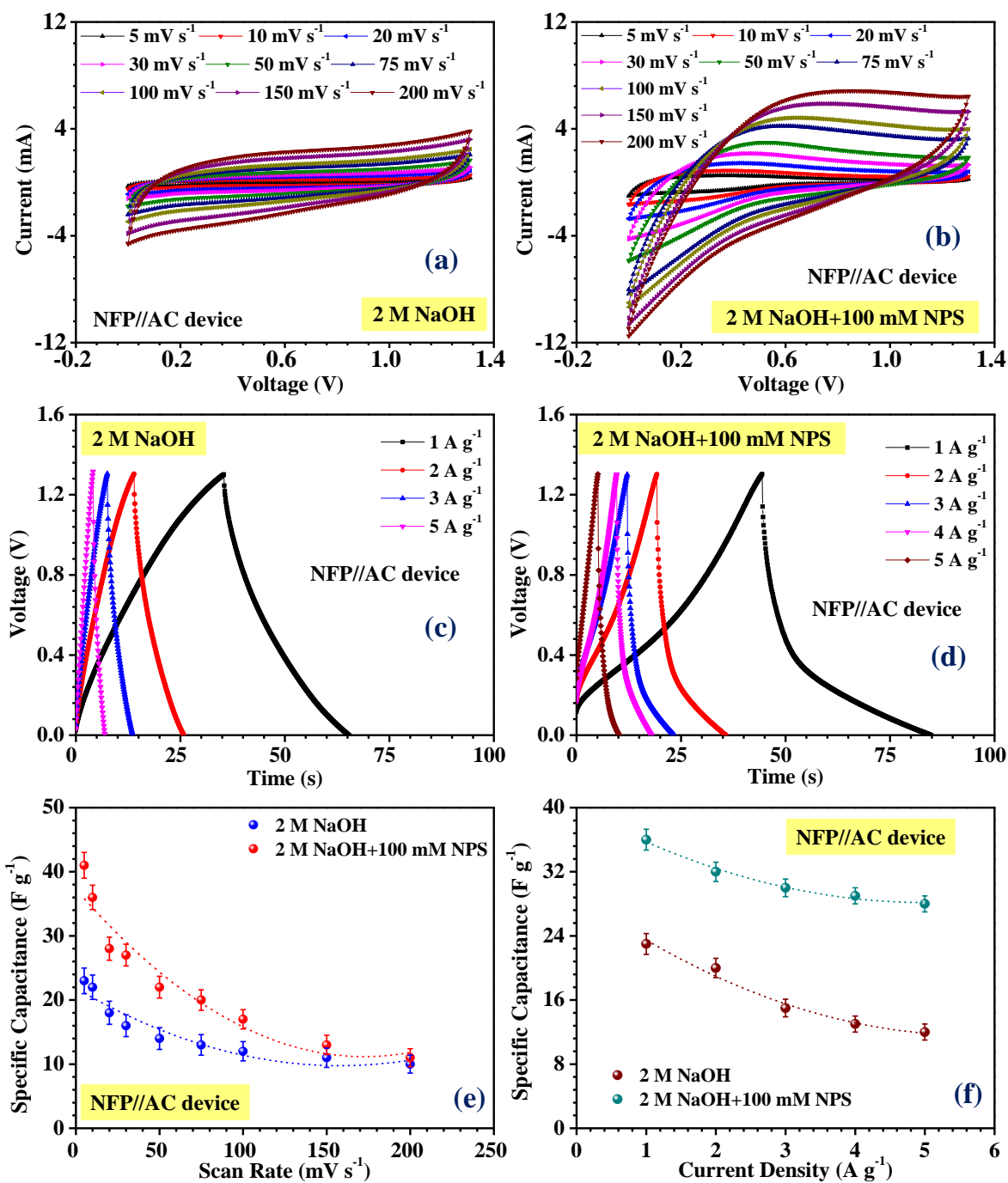

**Fig. 5** CV in (a) 2 M NaOH and (b) 2 M NaOH + 100 mM NPS; CD in (c) 2 M NaOH and (d) 2 M NaOH + 100 mM NPS; variation of specific capacitance in pure and redox modified electrolyte with (e) scan rate and (f) current density for NFP//AC device.

Fig. 6 shows the electrochemical behaviour of NFP//rGO asymmetric device in pure 2 M NaOH and 2 M NaOH + 100 mM NPS electrolyte. In pure electrolyte, the working potential window of the device was 1.7 V. The shape and size of the CV profile of the device assembled using pure NaOH showed relatively poor loop area [Fig. 6(a)]. In comparison, the redox modified device showed appreciable change in the CV profile. The loop area become wider confirming greater charge storge in the electrodes [Fig. 6(b)]. Along with the increment in the area of the CV curves, the small redox peaks were also visible, at low scan rates. The larger area of CV curves can be associated with the quasi reversible redox process. The additional charge storage could be linked to the higher number of redox reactions. The redox additive improvement of the capacitive behaviour of NFP//rGO device was further confirmed by the galvanostatic charge-discharge analysis. Fig. 6(c, d) depicts the charge-discharge profile of the device at various current densities in pure and 100 mM NPS added NaOH electrolyte, respectively. After redox additive addition, the CD profiles had a nonlinear character in the asymmetric device. This showed the enhancement in the redox activity of the device. The discharge time also increased, suggesting enhancement in the specific capacitance value along with an improved Coulombic efficiency of ~97%. The device in unmodified electrolyte delivered a maximum specific capacitance of ~ 55 and 63 F g$^{-1}$ at 5 mV s$^{-1}$ scan rate and 1 A g$^{-1}$ current density, respectively. The specific capacitance value of the device was estimated as ~88 and 99 F g$^{-1}$ at a scan rate and current density of 5 mV s$^{-1}$ and 1 A g$^{-1}$, respectively, after incorporation of NPS redox additive. Therefore, 44% increase in the specific capacitance for redox modified NFP//rGO device. At high current density of 5 A g$^{-1}$, the specific capacitance value decreased to ~45 F g$^{-1}$, indicating a rate capability of ~50%. Therefore, from the galvanostatic charge-discharge results, it could be inferred that the incorporation of NPS led to increment in the specific capacitance value of NFP//AC asymmetric device without any decrement in rate capability. This can be clearly seen from Fig. 6 (e, f).

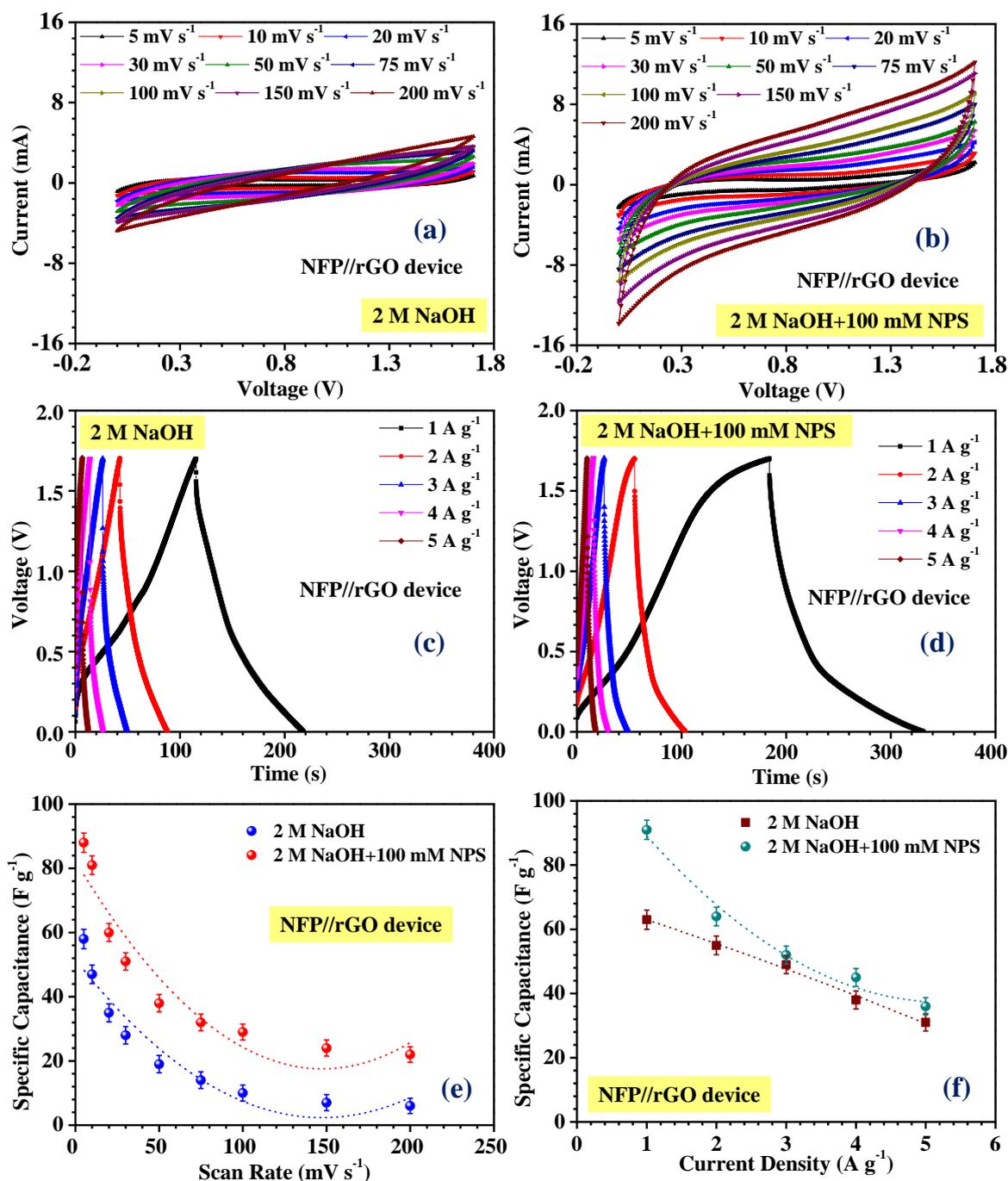

**Fig. 6** CV in (a) 2 M NaOH and (b) 2 M NaOH + 100 mM NPS; CD in (c) 2 M NaOH and (d) 2 M NaOH + 100 mM NPS; variation of specific capacitance in pure and redox modified electrolyte with (e) scan rate and (f) current density for NFP//rGO device.

To understand the electrochemical behaviour and charge storage mechanism of the device, electrochemical impedance spectroscopy (EIS) study of the fabricated device was performed, within the frequency range of 1 mHz-100 kHz, using small ac signal of 30 mV. In pure 2 M NaOH electrolyte, the impedance value showed a vertical rise in the low frequency region, indicating a capacitive nature of the fabricated device, as shown in Fig. 7(a, b). At the interface between electrode and electrolyte, the charge transfer resistance was estimated by the small semicircle formed at high frequency region, as shown in the inset of these figures. The diameter of the semicircle signifies the interfacial charge transfer resistance ($R_{ct}$) and the equivalent series resistance can also be calculated from the Nyquist plot, which is a combination of various types of resistances coming from electrode material, electrolyte and current collector along with the contact resistance.

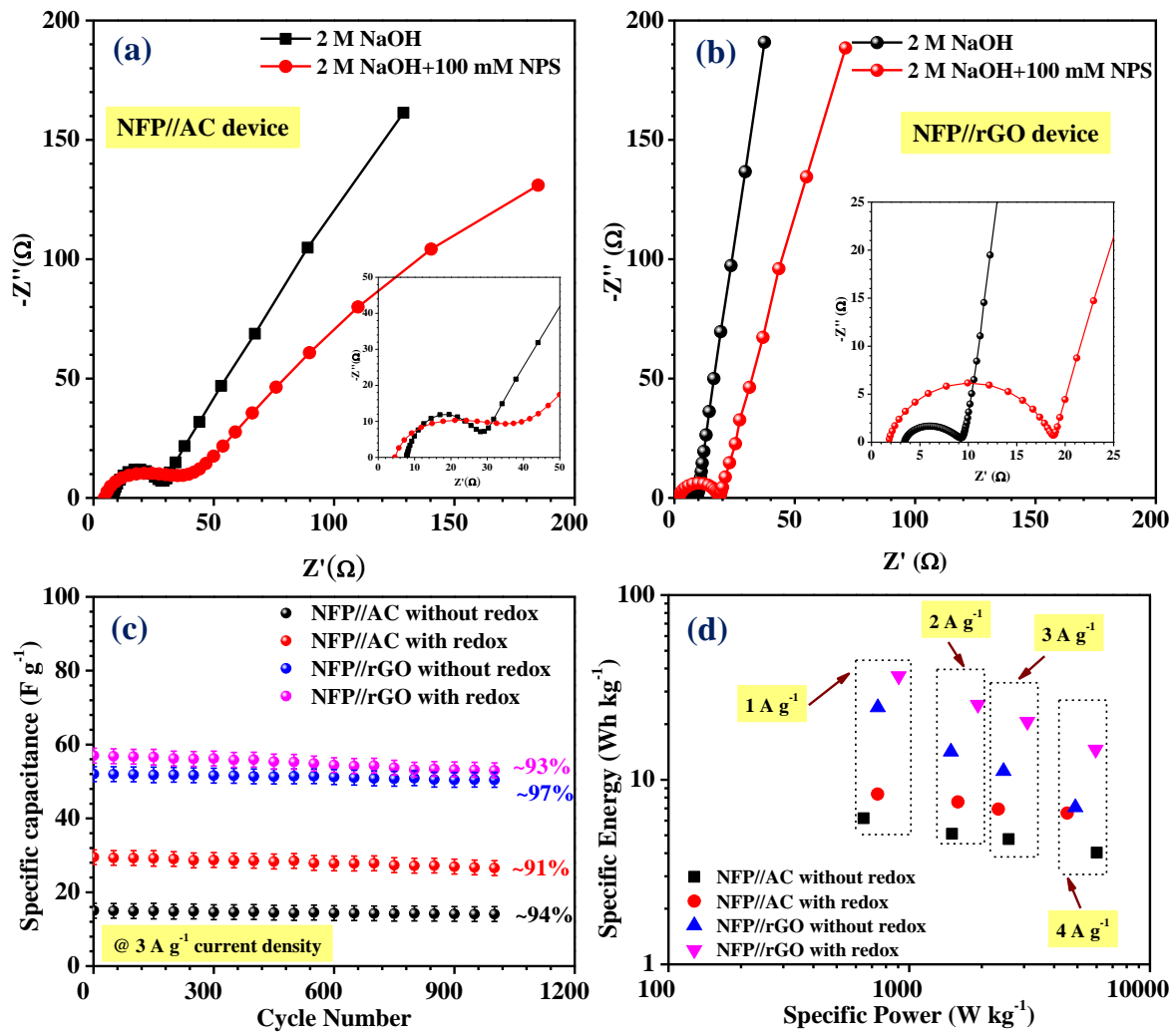

**Fig. 7** (a, b) Nyquist plots, (c) cycling stability and (d) Ragone plot of NFP//AC and NFP//rGO device, respectively, with and without redox additive.

In the pure and 100 mM NPS added NaOH electrolyte, the ESR values for NFP//AC device was found to be ~7.7 and 4.6 Ω, respectively. The ESR values of the NFP//rGO device was estimated to be ~3.3 and 1.9 Ω in pure and redox modified electrolyte, respectively. It could be seen that the ESR value decreased appreciably following redox additive addition, which can be attributed to the enhancement of the ionic conductivity of electrolyte by NPS. Additionally, NFP//rGO device showed lower resistance than the NFP//AC device. On the other hand, the charge transfer resistance was estimated as ~36 Ω, which was higher than the unmodified electrolyte that was 28 Ω for NFP//AC device. This increase in the charge transfer resistance indicated the restricted performance of the electrode because of the increasing electrolyte viscosity, which can obstruct the ion mobility. For the NFP//rGO device, the charge transfer resistance was found to be ~19 Ω after the addition of the redox additive compared to the unfodified electrolyte i.e. ~9 Ω.

To investigate the cycling stability, NFP//AC and NFP//rGO asymmetric device was subjected to repetitive charging and discharging upto 1000 cycles at 3 A g$^{-1}$ current density and the corresponding findings are shown in Fig. 7(c). In unmodified NaOH electrolyte, the NFP//AC device retained its specific capacitance value upto 1000 cycles by 94%. But, in redox-modified electrolyte, there was 91% capacity retention. The NFP//rGO device retained 97% of its specific capacitance value upto 1000 cycles. In redox-modified electrolyte 93% capacity retention was seen. NFP//rGO device showed better cycling stability compared to the NFP//AC device. Both the devices showed a decrement in the specific capacitance in the modified electrolyte, following cycling. The percenetage decrement in specific capacity, in both the devices were 3% after the 1000 cycling. The repetitive oxidation and reduction reaction at the electrode-electrolyte interface, because of the induced redox pairs, causes degradation of electrode material after a number of cycles, which can reduce the cycling stability of the device.

The energy and power density of a charge storage device is an indicator of its usefulness in real application. The specific energy ($E$) and power ($P$) of the device was calculated at various current densities using the following relations:

$$E = \frac{1}{2}CV^2 \quad \text{and} \quad P = \frac{E}{t}$$

where, $C$ is the specific capacitance at a particular current density, $V$ is the operating voltage window and $t$ is the total discharge time. The Ragone plot of the device fabricated using three different electrolytes is depicted in Fig. 7(d), for both the asymmetric devices. A maximum specific energy of ~6.18 W h kg$^{-1}$ with specific power of 647 W kg$^{-1}$ was delivered by NFP//AC asymmetric device assembled in 2 M NaOH electrolyte at room temperature. As expected, in redox modified electrolyte, the device showed improved performance by delivering higher energy density and specific power. The device could deliver the maximum energy density of ~36 Wh kg$^{-1}$ at a specific power of 905 W kg$^{-1}$, in NFP//rGO.

**Conclusion**

The hierarchical hollow structures of the NFP can be easily synthesized using a facile synthesis route. Competetive Na-ion supercapacitor can be realized with its proper combination with the redox active rGO. The specific capacitance was found to be 125 F g$^{-1}$ at a scan rate 5 mV s$^{-1}$ and 141 F g$^{-1}$ at a current density 1 A g$^{-1}$. rGO showed 196 F g$^{-1}$ specific capacitance with graphite current collector. Combining these two electrodes in a device can deliver 63 F g$^{-1}$ specific capacitance. To bring appreciable jump in this value so that the devie becomes invividually viable, the strategy of using redox additives is established. When a NPS modified NaOH electrolyte is used, nearly 50% enhancement is observed. The improvement in value is observed without compromising on cycling stability or coulumbic efficiency. Therefore, this establishes such Na-ion based supercapacitors as perfect alternative to more expensive Li-ion based systems.

**Acknowledgment**

The author acknowledges the financial support received from DST (India) under the MES scheme to pursue work under the project entitled: "Hierarchically nanostructures energy materials for next-generation Na-ion based energy storage technologies and their use in renewable energy systems" [Grant number: DST/TMD/MES/2k16/77].

# Supplementary Information

# Role of redox additive modified electrolytes in making Na-ion supercapacitors a competitive energy storage device

Sudipta Biswas[a], Debabrata Mandal[b], Ananya Chowdhury[a] and Amreesh Chandra[a,b*]

[a]Department of Physics, [b]School of Nano Science and Technology,

Indian Institute of Technology Kharagpur, Kharagpur-721302.

Email: *achandra@phy.iitkgp.ac.in*

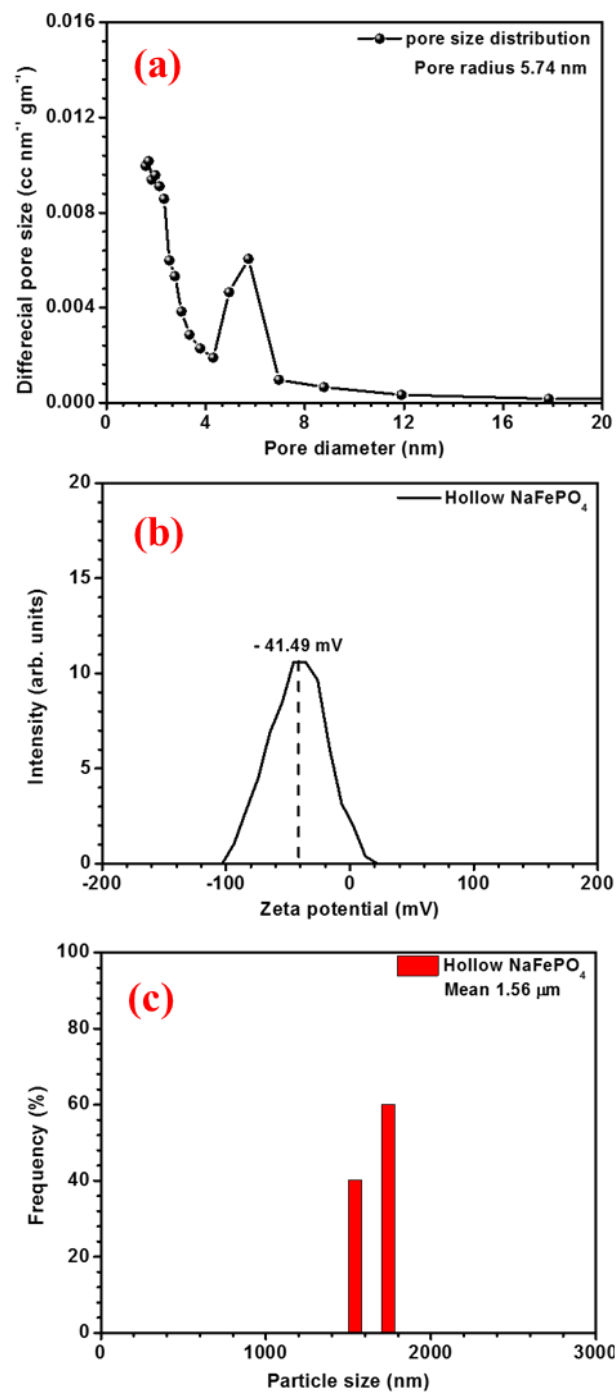

**Fig. S1** (a) pore size distribution, (b) zeta potential and (c) particle size distribution for NaFePO$_4$.

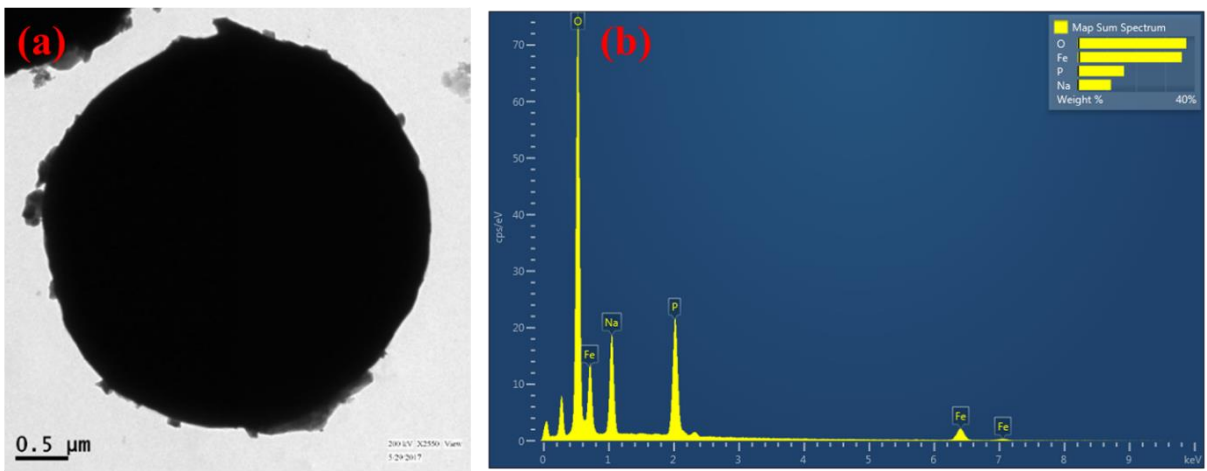

**Fig. S2** (a) TEM micrograph and (b) EDAX spectra of the NaFePO$_4$ structures.

X-ray photoelectron spectroscopy (XPS) measurement is shown in Fig. 1(b-f). The complete spectra of XPS shows peaks of Na, Fe, P, O and C elements without any other impurities. The Fe 2p spectrum (Fig. 2(d)) with two characteristic peaks at 710.5 eV (Fe $2p^{3/2}$) and 724.2 eV (Fe $2p^{1/2}$) indicates the bivalence $Fe^{2+}$.

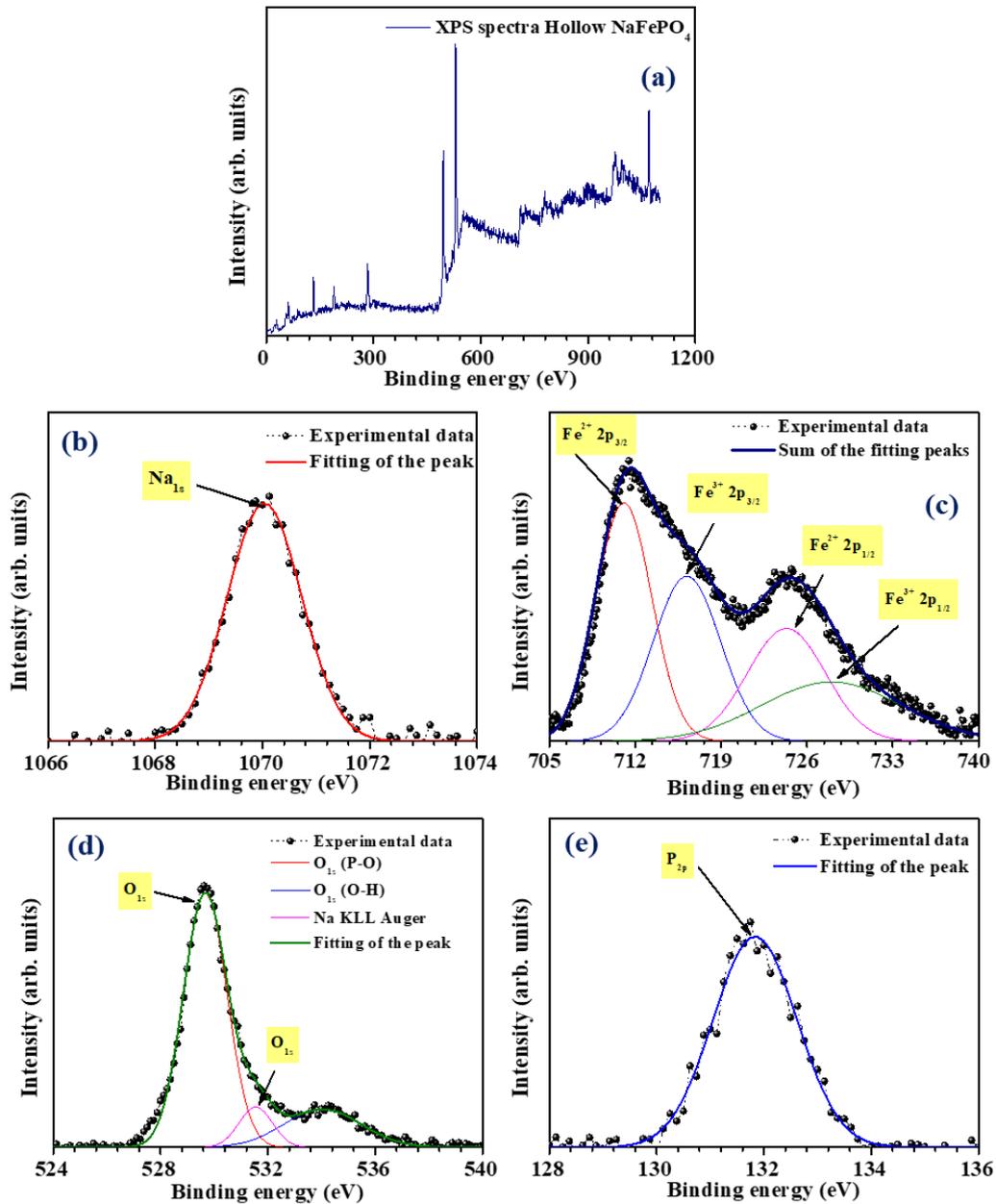

**Fig. S3** (a) XRD spectra of carbon and RGO.

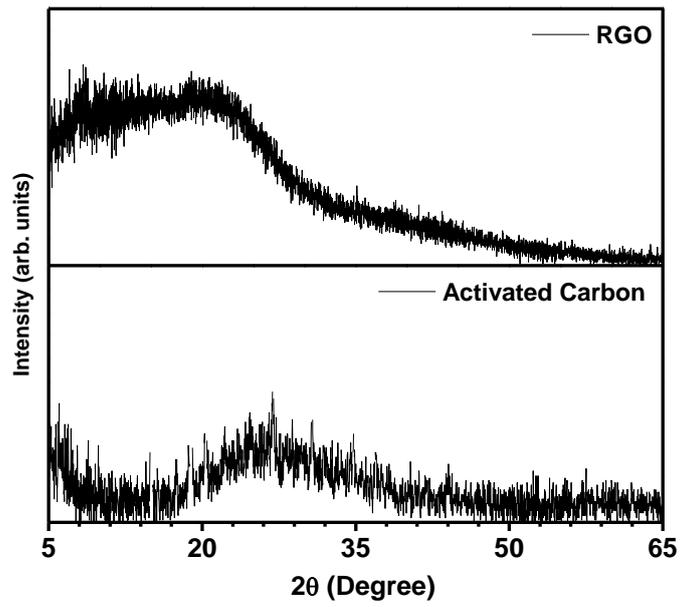

**Fig. S4** XRD spectra of carbon and rGO.

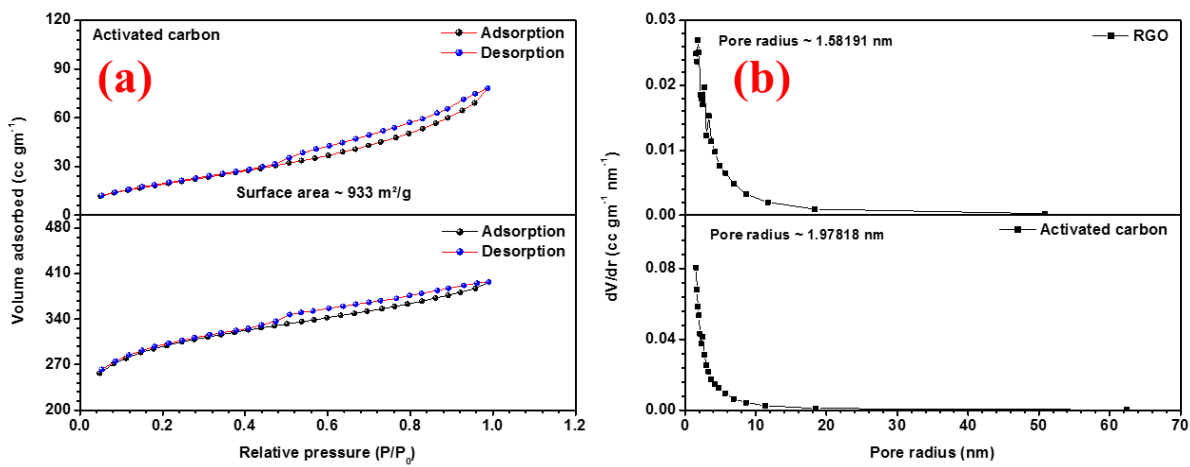

**Fig. S5** (a) N$_2$ adsorption and desorption curve and (b) pore size distribution of activated carbon and rGO, respectively.

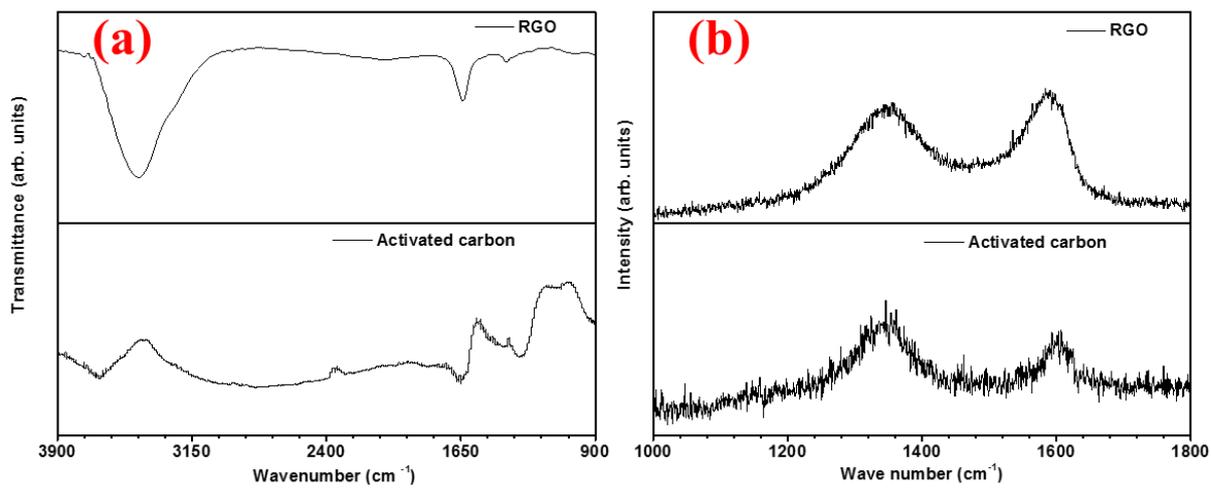

**Fig. S6** (a) FTIR spectra and (b) Raman spectra of activated carbon and rGO.

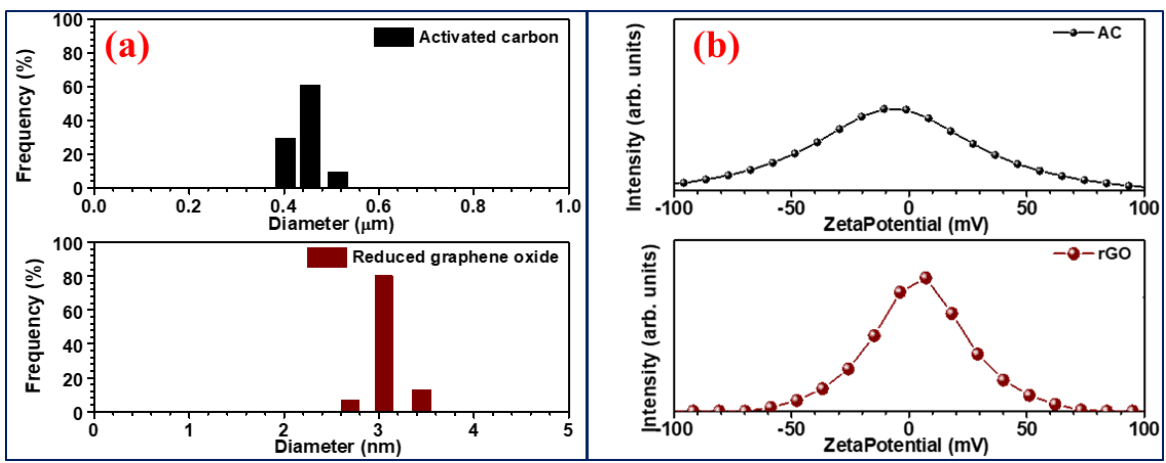

**Fig. S7** (a) Particle size distribution and (b) zeta potential activated carbon and rGO.

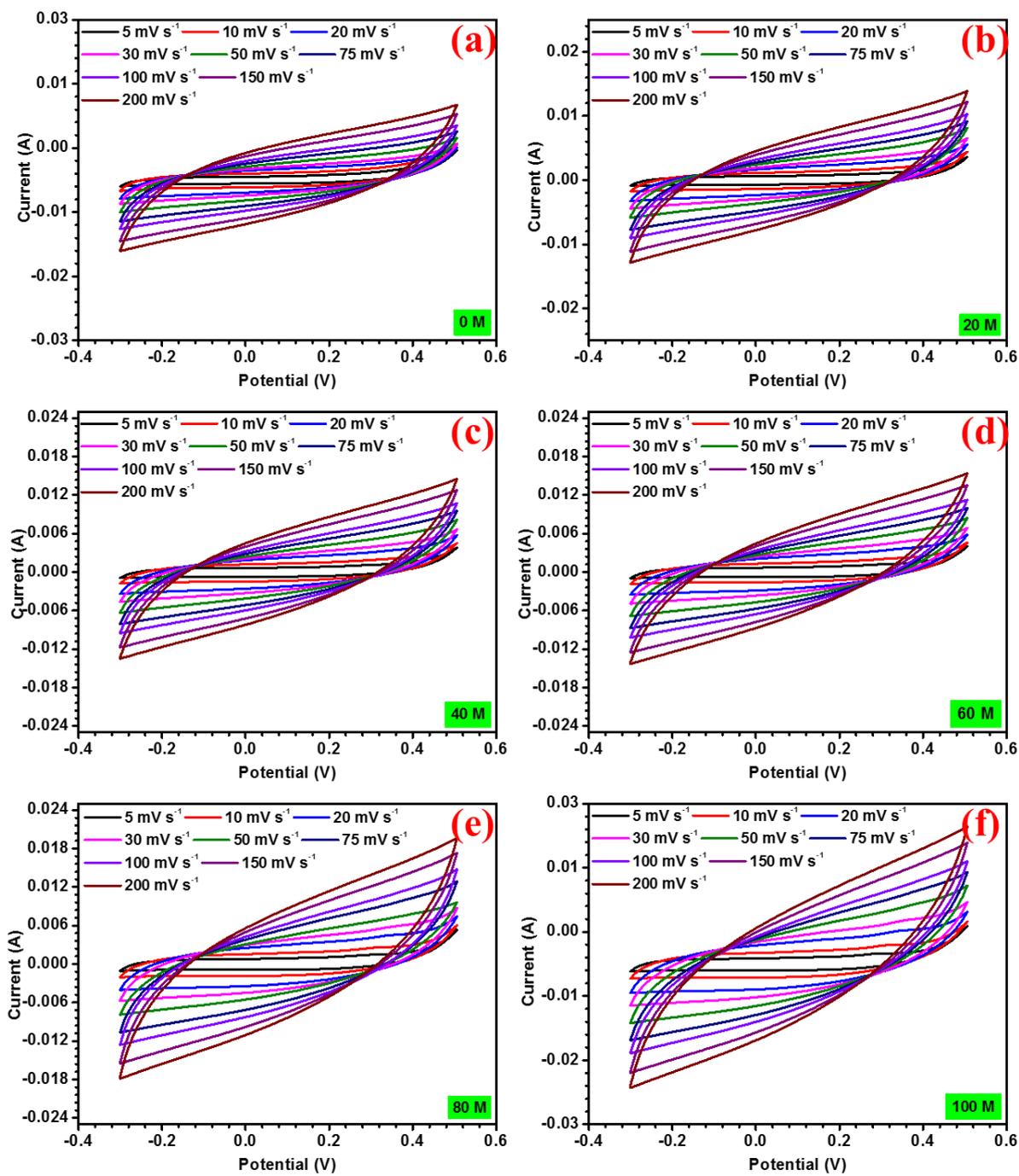

**Fig. S8** (a-f) CV of NaFePO$_4$ with increasing concentration of NPS (0 mM to 100 mM).

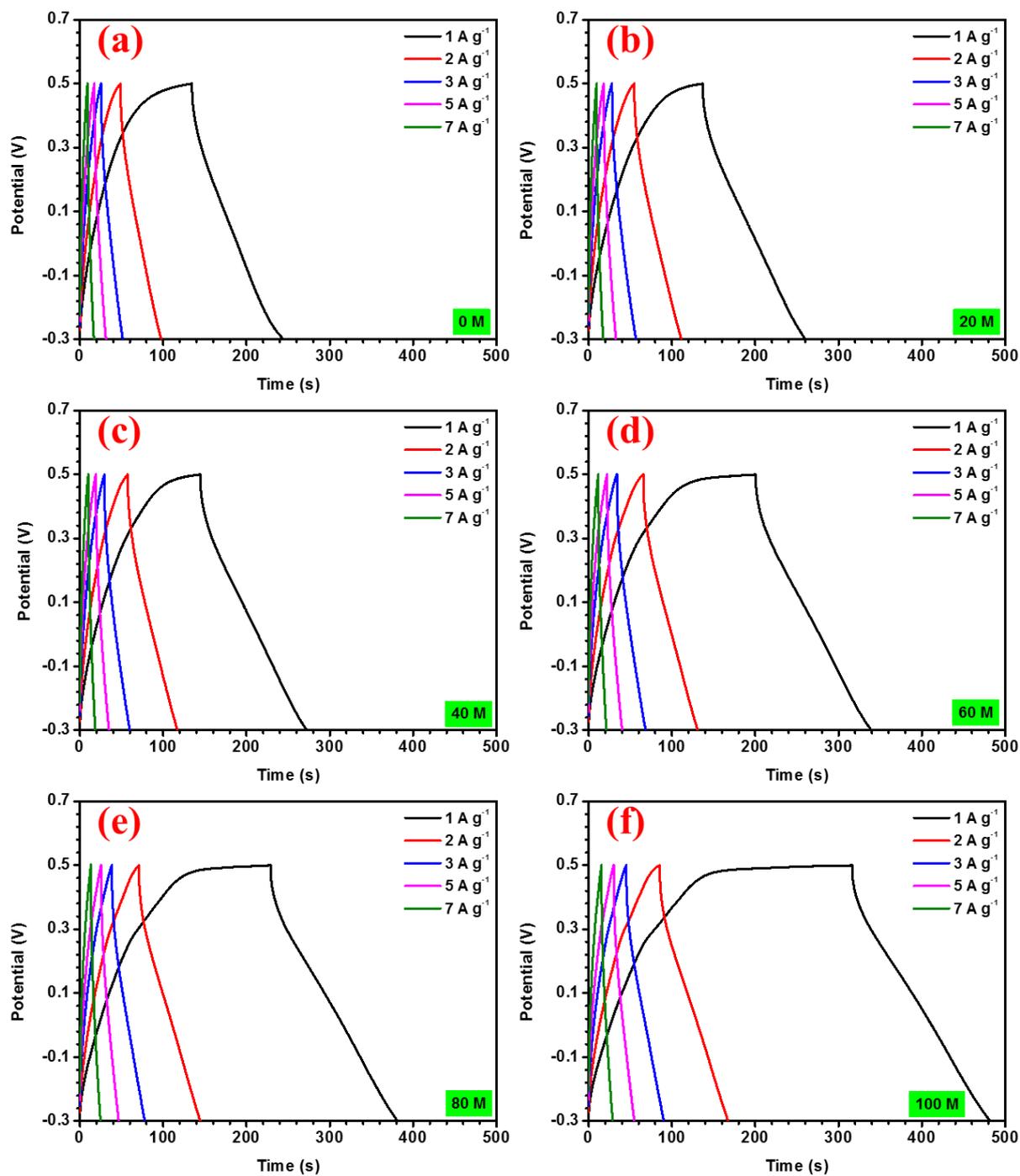

**Fig. S9** (a-f) CD of NaFePO$_4$ with increasing concentration of NPS (0 mM to 100 mM).

In this work, asymmetric devices have been fabricated using NaFePO$_4$ as the positive and carbon materials as the negative electrode. To fabricate the device, charge balance was done using the following mass balance relation:

$$\frac{m_+}{m_-} = \frac{V_- C_-}{V_+ C_+} \quad \text{(S1)}$$

where, C$_-$ and C$_+$ are the capacitances (in F g$^{-1}$) measured at the same scan rate (usually 5 mV s$^{-1}$) i.e., the lowest scan rate, using the three-electrode system, while V$_+$ and V$_-$ denote the working potential window for the positive and negative electrodes, respectively.

The average specific capacitance from the CV curves is calculated using the relation:

$$C_{CV} = \frac{1}{2m\nu \triangle V} \int_{-V}^{+V} I \, dV \quad \text{(S2)}$$

where m, $\nu$, $\Delta V$ and $\int_{-V}^{+V} I \, dV$ denote the mass of the active material, scan rate, potential window, and absolute area under the CV curve, respectively.

The capacity to deliver the power of commercial a supercapacitor is obtained by the values of specific capacitance measured in galvanostatic charge-discharge (CD) curves. The specific capacitance values are calculated using the relation

$$C_{CD} = \frac{I \, dt}{m \, (V - IR)} \quad \text{(S3)}$$

where, I/m, dt, V and IR denotes the current density, time of discharge, voltage window and the voltage drop during discharge, respectively.

The specific capacitance of the device was estimated using the following relations:

$$C_{CV} = \frac{1}{m\nu \triangle V} \int_{-V}^{+V} I \, dV \quad \text{(for CV)} \quad \text{(S4)}$$

$$C_{CD} = \frac{I \, dt}{m \, (V - IR)} \quad \text{(for CD)} \quad \text{(S5)}$$

where, '$m$', '$\nu$', '$\triangle V$' and '$\int_{-V}^{+V} I \, dV$', '$I$', '$dt$' and '$IR$' denote total mass of positive and negative electrode, scan rate, working potential window, area of the CV profile, discharging current, discharging time and the potential drop during discharge respectively.

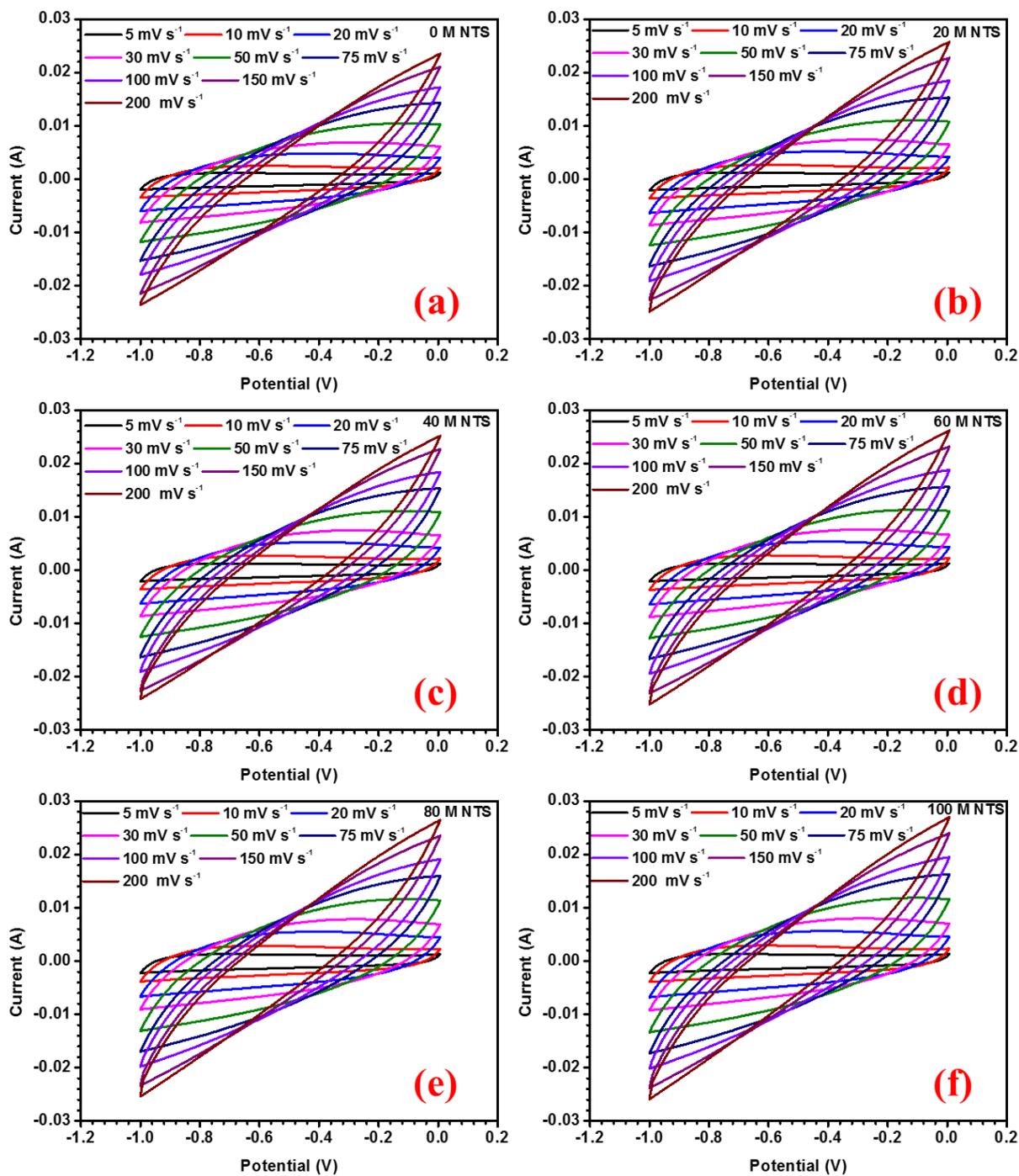

**Fig. S10** (a-f) CV of activated carbon with increasing concentration of NPS (0 mM to 100 mM).

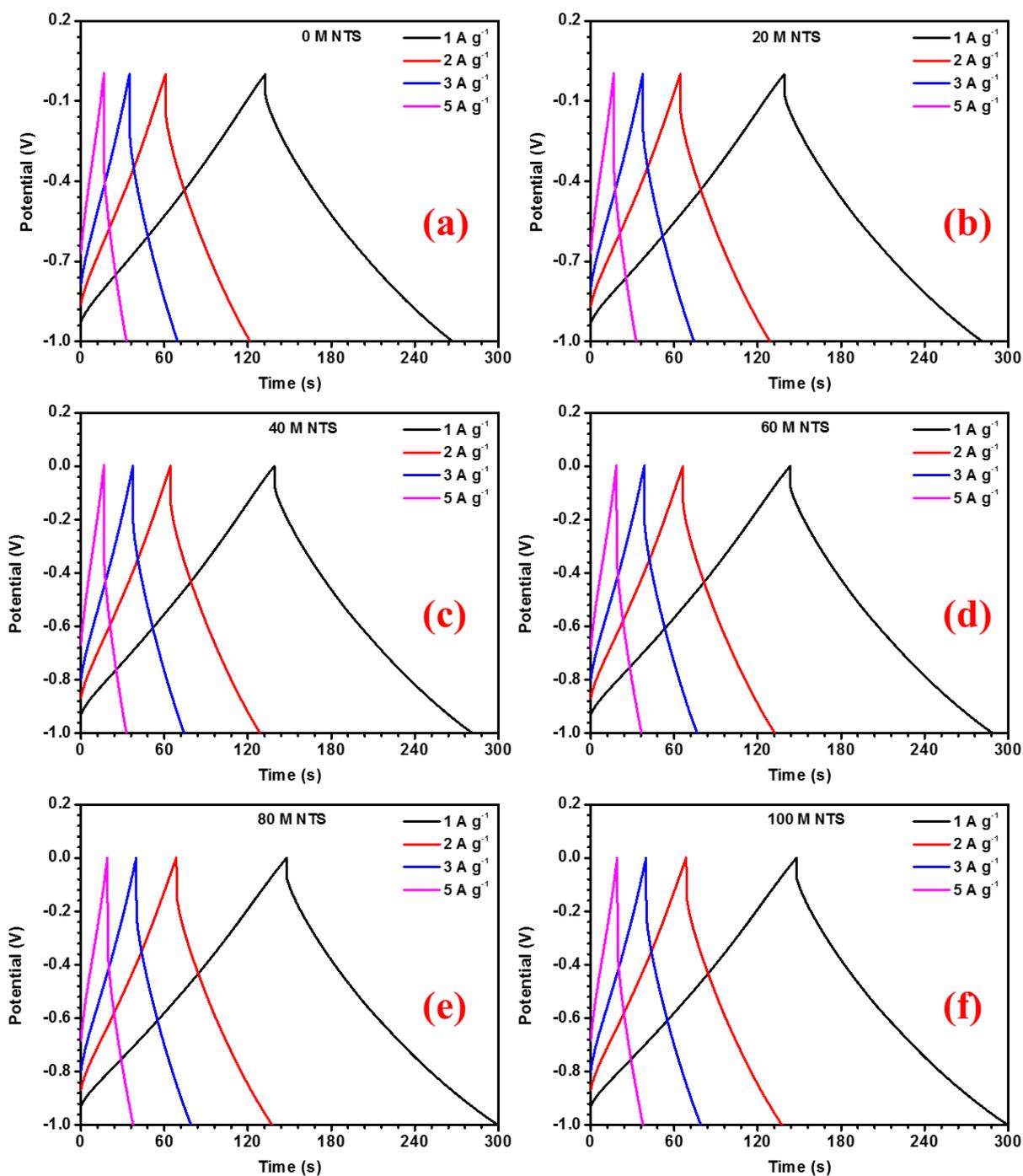

**Fig. S11** (a-f) CD of activated carbon with increasing concentration of NPS (0 mM to 100 mM).

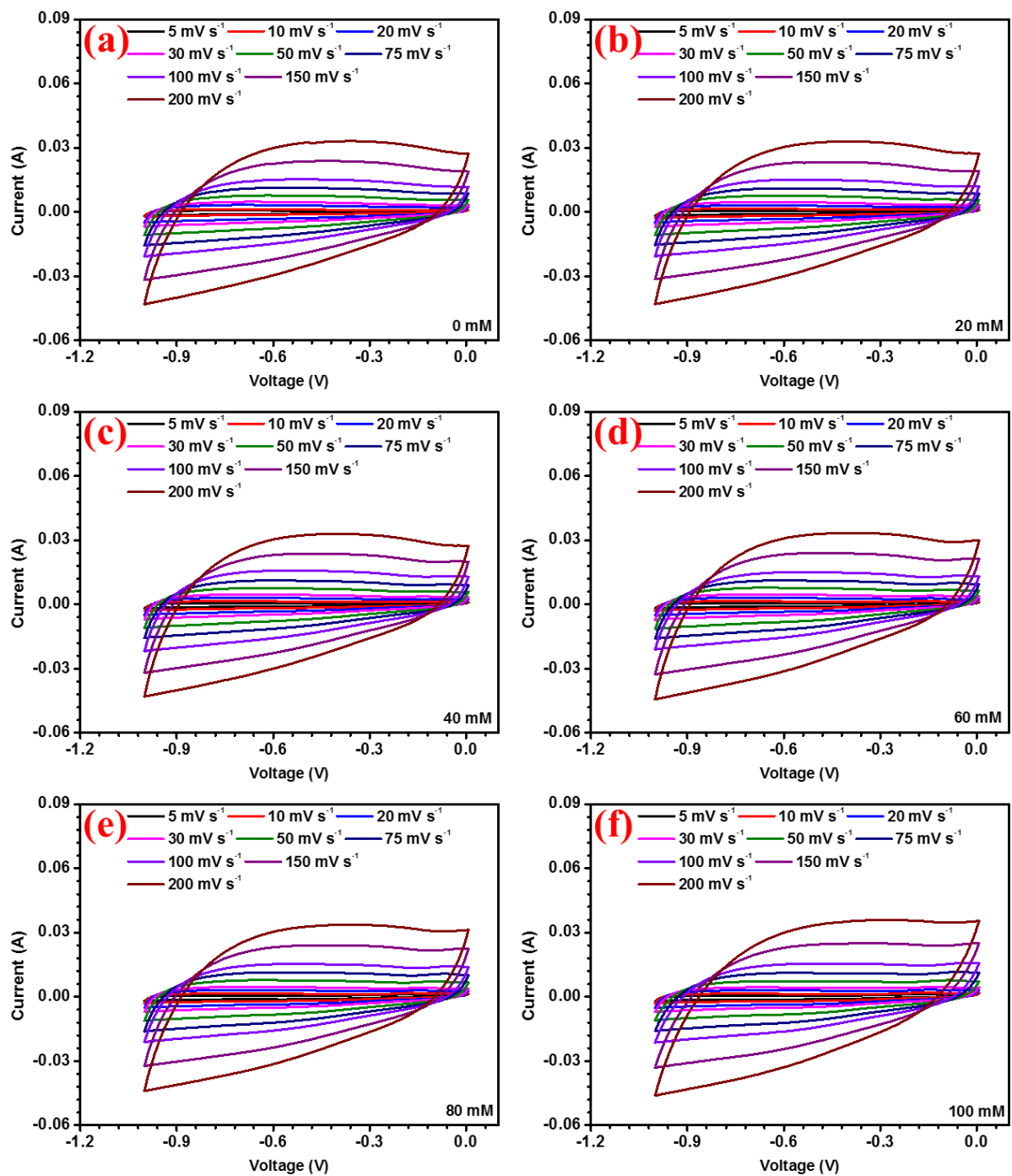

**Fig. S12** (a-f) CV of rGO with increasing concentration of NPS (0 mM to 100 mM).

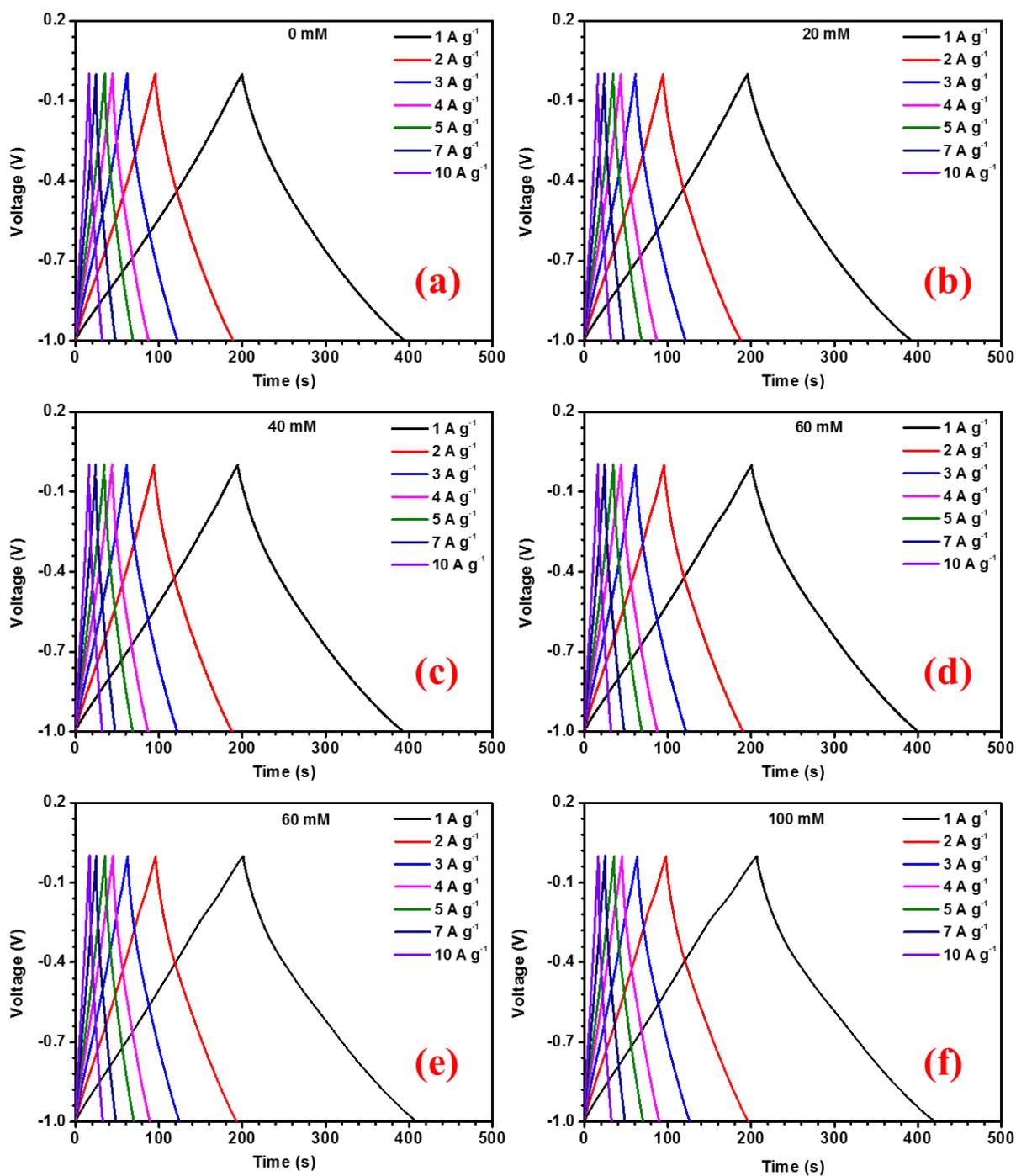

**Fig. S13** (a-f) CV of rGO with increasing concentration of NPS (0 mM to 100 mM).

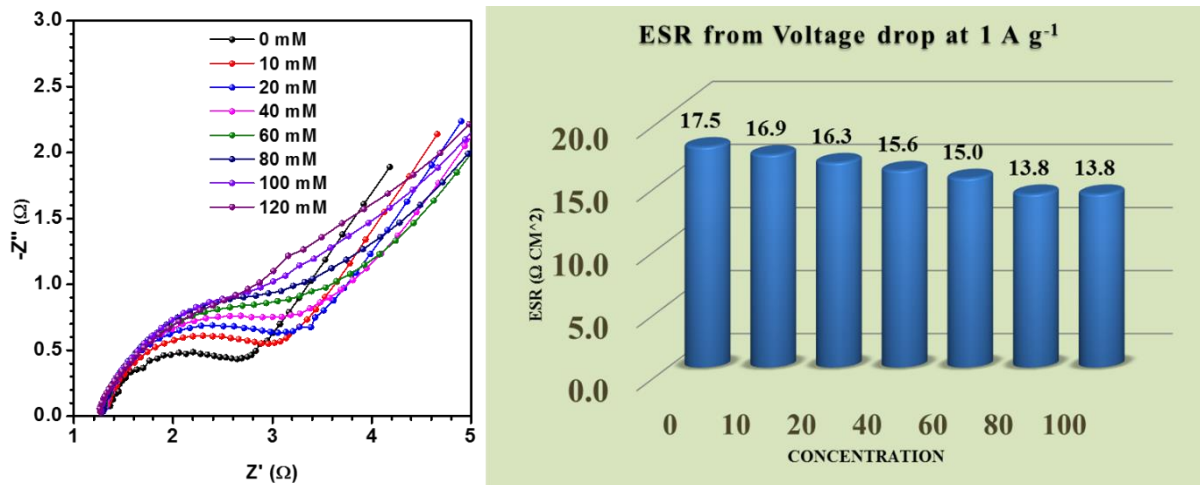

**Fig. S14** ESR of NaFePO₄ electrode with increasing concentration of NPS (0 mM to 100 mM).

**Table** S1 Specific capacitance of NaFePO$_4$ with increasing scan rate in 2 M NaOH with increasing concentration of NPS redox additive.

| Scan rate (mV s$^{-1}$) | Specific Capacitance (F g$^{-1}$) | | | | | | |
|---|---|---|---|---|---|---|---|
| | 0 mM | 10 mM | 20 mM | 40 mM | 60 mM | 80 mM | 100 mM |
| 5 | 125 | 134 | 141 | 148 | 158 | 178 | 194 |
| 10 | 114 | 118 | 127 | 137 | 148 | 173 | 192 |
| 20 | 87 | 96 | 100 | 110 | 122 | 148 | 168 |
| 30 | 70 | 77 | 82 | 90 | 100 | 123 | 140 |
| 50 | 53 | 59 | 62 | 66 | 73 | 85 | 102 |
| 75 | 44 | 50 | 51 | 54 | 58 | 74 | 80 |
| 100 | 39 | 43 | 44 | 46 | 49 | 64 | 74 |
| 150 | 32 | 36 | 36 | 37 | 39 | 49 | 64 |
| 200 | 31 | 30 | 30 | 31 | 32 | 42 | 52 |

**Table** S2 Specific capacitance of activated carbon with increasing scan rate in 2 M NaOH with increasing concentration of NPS redox additive.

| Scan rate (mV s$^{-1}$) | Specific Capacitance (F g$^{-1}$) | | | | | |
|---|---|---|---|---|---|---|
| | 0 mM | 20 mM | 40 mM | 60 mM | 80 mM | 100 mM |
| 5 | 158 | 165 | 166 | 169 | 174 | 178 |
| 10 | 152 | 161 | 161 | 165 | 170 | 173 |
| 20 | 131 | 141 | 141 | 144 | 147 | 150 |
| 30 | 113 | 121 | 121 | 123 | 125 | 128 |
| 50 | 84 | 88 | 88 | 90 | 91 | 93 |
| 75 | 60 | 64 | 64 | 65 | 64 | 62 |
| 100 | 45 | 48 | 48 | 48 | 48 | 49 |
| 150 | 28 | 30 | 30 | 30 | 30 | 31 |
| 200 | 19 | 21 | 21 | 21 | 21 | 21 |

| Current density (A $g^{-1}$) | Specific Capacitance (F $g^{-1}$) | | | | | |
|---|---|---|---|---|---|---|
| | 0 mM | 20 mM | 40 mM | 60 mM | 80 mM | 100 mM |
| 1 | 1 | 144 | 152 | 153 | 155 | 162 |
| 2 | 2 | 143 | 148 | 148 | 154 | 158 |
| 3 | 3 | 131 | 140 | 144 | 144 | 145 |
| 5 | 5 | 125 | 125 | 126 | 133 | 142 |

**Table** S3 Specific capacitance of activated carbon with increasing current density in 2 M NaOH with increasing concentration of NPS redox additive.

| Scan rate (mV s$^{-1}$) | Specific Capacitance (F g$^{-1}$) | | | | | |
|---|---|---|---|---|---|---|
| | 0 mM | 20 mM | 40 mM | 60 mM | 80 mM | 100 mM |
| 5 | 188 | 189 | 189 | 196 | 209 | 213 |
| 10 | 186 | 187 | 187 | 195 | 203 | 206 |
| 20 | 180 | 180 | 180 | 186 | 191 | 193 |
| 30 | 175 | 176 | 179 | 182 | 183 | 183 |
| 50 | 168 | 170 | 171 | 177 | 179 | 179 |
| 75 | 163 | 164 | 164 | 167 | 169 | 172 |
| 100 | 160 | 162 | 162 | 163 | 167 | 166 |
| 150 | 158 | 158 | 161 | 162 | 164 | 163 |
| 200 | 156 | 158 | 158 | 161 | 161 | 162 |

**Table** S4 Specific capacitance of rGO with increasing scan rate in 2 M NaOH with increasing concentration of NPS redox additive.

Table S5 Specific capacitance of rGO with increasing current density in 2 M NaOH with increasing concentration of NPS redox additive.

| Current density (A g$^{-1}$) | Specific Capacitance (F g$^{-1}$) | | | | | | |
|---|---|---|---|---|---|---|---|
| | 0 mM | 10 mM | 20 mM | 40 mM | 60 mM | 80 mM | 100 mM |
| 1 | 196 | 196 | 197 | 198 | 201 | 209 | 215 |
| 2 | 189 | 189 | 189 | 190 | 195 | 196 | 201 |
| 3 | 182 | 184 | 185 | 186 | 187 | 193 | 194 |
| 4 | 173 | 175 | 174 | 176 | 180 | 182 | 187 |
| 5 | 167 | 169 | 170 | 170 | 175 | 179 | 182 |
| 7 | 158 | 161 | 162 | 164 | 166 | 171 | 171 |
| 10 | 153 | 158 | 159 | 159 | 159 | 164 | 167 |